\documentclass[12pt]{article}
\usepackage{graphicx}
\usepackage{amsmath}
\usepackage{amssymb}
\usepackage{url}

\usepackage[latin1]{inputenc}
\usepackage[T1]{fontenc}
\usepackage[english]{babel}

\usepackage{bm}

\title{Polarization in Low Frequency Radio Astronomy}
\author{Baptiste Cecconi}
\date{2016}

\begin{document}

\maketitle

\section*{Editorial Notice}
This content of this document has been initially written in 2016 as a chapter of a book entitled ``The Universe in polarised light'', after the ``Ecole Internationale de Polarim\'etrie en Astrophysique'' (International School on Polarimetry for Astrophysics) organized in June 2013, at Centre Paul-Langevin, Aussois, France. This summer school has been funded by the EU funded COST action MP1104\footnote{\url{http://www.polarisation.eu}} and OSUG (Observatoire des Science de l'Univers de Grenoble). This book has not been published yet due to delays and lack of funding.

\section*{Foreword}
This chapter introduces the concepts of polarimetry in the case of low frequency radio astronomy. In this regime radio waves are usually not the signature of atomic or molecular transitions lines, but rather that of unstable particle distribution functions releasing their free energy through electromagnetic radiation. As the radio source region is usually magnetized, the propagation medium (at least close to the source) is anisotropic, and the polarization depends on the local magnetic field direction, the propagation mode and the direction of propagation.

\section{Introduction}
Low frequency radio astronomy is defined by a frequency span ranging from $\sim$1\,kHz to $\sim$100\,MHz (or $\sim$3\,m to $\sim$300\,km in terms of wavelengths). The energy of a single photon at 100\,kHz is 7$\times$10$^{-29}$\,J (or 4$\times$10$^{-10}$\,eV). The wave properties of light are used instead of its corpuscular properties. The electric (and/or magnetic) field fluctuations of the wave are sampled. The polarization of the radio waves is thus directly measured by the sensors. Depending on the sensor physical shape different polarization components are intrinsically measured: linear electric dipoles measure linear polarization, loop or helicoidal antenna are measuring the circular polarization. 

\begin{figure}
\centering\includegraphics[width=0.8\textwidth]{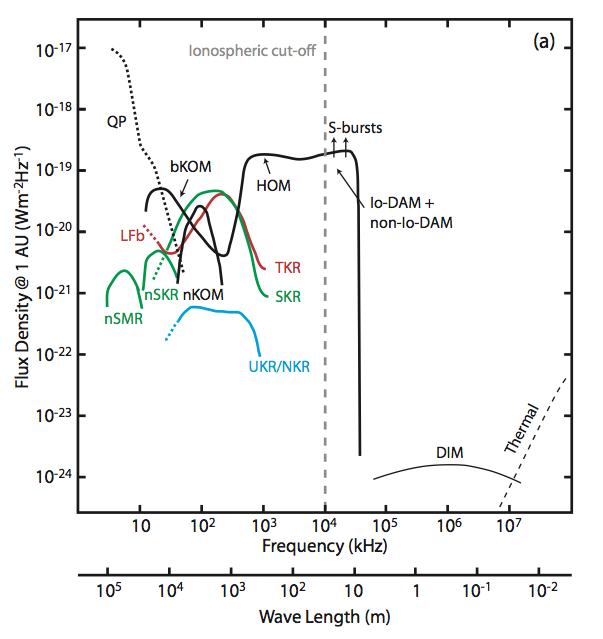}
\caption{Compared planetary radio emission spectral flux densities. The radio emissions from Jupiter are in black (\emph{nKOM}: narrowband Kilometric Radiation; \emph{bKOM}: broadband Kilometric Radiation; \emph{QP}: Quasi-Periodic Bursts; \emph{HOM}: Hectometric Radiation; \emph{S-bursts}: Short (or Millisecond) Bursts; \emph{Io-DAM}: Io-controlled Decametric Radiation; \emph{non-Io-DAM}: non-Io-controlled Decametric Radiation; \emph{DIM}: Decimetric Radiation; \emph{Thermal}: Black body thermal radiation), from Saturn in green (\emph{n-SMR}: narrowband Saturn Myriametric Radiation; \emph{n-SKR}: narrowband Saturn Kilometric Radiation; \emph{SKR}: Saturn Kilometric Radiation), from Earth in red (\emph{LFb}: Low-Frequency Bursts; \emph{TKR} (or \emph{AKR}): Terrestrial (or Auroral) Kilometric Radiation), and from Uranus and Neptune in blue (\emph{UKR}/\emph{NKR}: Uranus/Neptune Kilometric Radiation). The vertical grey dashed line represents the Earth ionospheric cut-off. Figure Adapted from \cite{zarka_1992}, updated with \cite{zarka_2004,lamy_2008,lamy_2010}.}\label{fig:1}
\end{figure}

The electromagnetic radiations that can be detected in the low frequency range have various origins. In the vicinity of the Earth, the most intense sources are the Sun and the magnetized planets, as shown in Figure (\ref{fig:1}). Human technologies are strong sources of Radio Frequency Interferences (RFI). Atmospheric electricity (i.e., Earth and planetary lightnings) emit wide band electromagnetic pulses. Our Galaxy is radiating a continuum of emission (called the Galactic Background), resulting from free-free interactions of electrons in the interstellar medium. The Galactic Background is almost isotropic and its brightness temperature exceeds 10$^7$\,K at 1\,MHz \cite{cane_79,dulk_2001,manning_2001}. The brightness temperature of Solar radio bursts and of Jovian radio emissions are respectively of the order of $\sim$10$^{12}$\,K and $\sim$10$^{18}$\,K. Those brightness temperatures usually do not correspond to the actual black body temperature of the radio source medium. Hence, these radio emissions are called ``non-thermal'' radio emissions.

Solar and Solar Wind radio sources are produced by populations of relativistic electrons in the Solar Corona or escaping the Sun in the interplanetary medium. The two main low frequency Solar radio sources are called type II and type III solar radio bursts. Type II bursts are emitted by electrons accelerated in front of coronal mass ejections and interplanetary shocks (see \cite{gopal_2009} and references therein). Type III bursts are produced by beams of electrons ejected from the Sun and travelling outward along magnetic field lines. The radio waves are believed to be produced by wave conversion from strong Langmuir waves excited along the beam path (see \cite{henri_2009,reid_2014} and references therein). Radio waves are scattered in the inner heliosphere by solar wind inhomogeneities, resulting in an apparent spatially extended source (up to 90$^\circ$ at $\sim$100\,kHz \cite{steinberg_85}).

In planetary magnetospheres, the two main low frequency radio emissions families are the auroral radio emissions and the radiation belts. Planetary aurorae result from particle precipitations towards the magnetic poles of the planet. When those particles reach the planetary ionosphere, they transfer their kinetic energy through collisions with the atmospheric populations of atoms and molecules, which are releasing this energy by emitting photons in the infrared to ultraviolet range, depending on the chemical species present in the medium. A fraction of the particles does not reach the planetary ionosphere thanks to the magnetic mirror effect. The up-going particles are the source of the auroral radio emissions. The typical energy of particles responsible for  auroral radio emissions is of the order of 10\,keV. The electrons are the main driver for electromagnetic auroral phenomena and the emission mechanism is the Cyclotron MASER Instability (CMI), see \cite{louarn_1992,zarka_1992,treuman_2000} for more details. This non-thermal process has been observed in-situ at Earth by various instruments onboard the Viking \cite{hilgers_92} and FAST \cite{ergun_98} space missions. Remote observations at Jupiter and Saturn are consistent with the CMI phenomenology \cite{zarka_1992,hess_2008,cecconi_2009}. Furthermore, radio observations by Cassini near a SKR radio source are showing results that are fully consistent with the cold plasma theory and the CMI emission process \cite{lamy_2010,lamy_2011}. The apparent brightness temperature of such radio source can be higher than 10$^{15}$\,K. The radio waves are emitted at a frequency close to the local cyclotron frequency. The radio source is strongly anisotropic. Its theoretical beaming pattern is a hollow cone, aligned with the ambiant magnetic field, with an opening half angle between 45$^\circ$ and 90$^\circ$, and a thickness of $\sim$1$^\circ$. 

The planetary radiation belts radio emissions are produced by synchrotron emission of highly relativistic electrons trapped in the planetary magnetic field near the magnetic equator. The radio source is the radiation belts by itself and spans up to several planetary radii in the equatorial plane. The Jovian radiation belts are the most intense of the kind in the solar system, and span from about 40\,MHz up to $\sim$10\,GHz \cite{zarka_1992,santoscosta_08,girard_2016}. The terrestrial radiation belts are emitting a radio emission around 1\,MHz, but have still been poorly studied. 

Astrophysical objects such as pulsars are also emitting in this frequency range. Magnetized exoplanets are believed to host low frequency radio sources, similarly the Solar System planets \cite{zarka_2007,hess_2011,nichols_2012,zarka_2012}. Finally, the Dark Ages and Cosmic Dawn radiation signatures is also predicted to appear in this frequency range (see \cite{burns_11} and references therein).

\section{Magneto-ionic theory}
This section presents the theoretical background required to understand the emission and propagation of low frequency radio emissions, and thus their polarization.

\subsection{Presentation of the assumptions and conditions}
A series of preliminary assumptions are made: 
\begin{itemize}
\item The propagation medium is a plasma, which is globally neutral.
\item The plasma is cold, which implies: (i) without perturbations, the particles are not moving; (ii) particles have no thermal velocity; (iii) the pressure gradient is negligible.
\item The effect of the ions is negligible, as the radio frequency range is much higher than that of the ion scales.
\item The medium is magnetized and thus anisotropic.
\end{itemize}
\subsection{Basic Equations}
Using the Fourier notation, the Maxwell-Faraday and Maxwell-Amp\`ere laws respectively write:
\begin{align}
i\bm{k}\times\bm{E}&=i\omega\bm{B}\\
i\bm{k}\times\bm{B}&=-i\frac{\omega}{c^2}\overline{\overline{\bm{\kappa}}}.\bm{B}
\end{align}
The equation driving the electric field can thus be derived as:
\begin{align}
\bm{n}\times(\bm{n}\times\bm{E})+\overline{\overline{\bm{\kappa}}}.\bm{E}=0
\end{align}
using the refractive index $\bm{n}$ vector defined as:
\begin{align}
\bm{n}=\frac{n}{k}\bm{k}=\frac{c}{\omega}\bm{k}
\end{align}
The permittivity tensor is computed from the equation of motion of the particles in the plasma. After the assumptions presented in the previous paragraph, it takes the following form:
\begin{align}
\bm{\kappa}=\left(\begin{array}{ccc}
S&-iD&0\\
iD&S&0\\
0&0&P
\end{array}\right)
\end{align}
with $S=1-X/(1-Y^2)$, $D=XY/(1-Y^2)$, $P=1-X$, and defining $X=(\omega_p/\omega)^2$ and $Y=\omega_c/\omega$. The characteristic frequencies $\omega_p$ and $\omega_c$ are respectively the plasma frequency and the electron cyclotron frequency. 

\subsection{Refraction index: Appleton-Hartree equation}
As the medium is magnetized, we choose to represent the various vectors in the reference frame defined as: $\bm{z}$ is along the main ambiant magnetic field direction $\bm{B}_0$, and the wave vector $\bm{k}$ is in the $(\bm{x},\bm{z})$ plane. Hence the refractive index vector is $\bm{n}=(n\sin\theta,0,n\cos\theta)$,
with $\theta$ the angular separation between $\bm{B}_0$ and $\bm{k}$. 

The dispersion equation derives from the equation of the electric field of the wave, by searching for the eigenmodes of propagation, i.e., solving the following linear system:
\begin{align}
\left(\begin{array}{ccc}
S-n^2\cos^2\theta&-iD&n^2\sin\theta\cos\theta\\
iD&S-n^2&0\\
n^2\sin\theta\cos\theta&0&P-n^2\sin^2\theta
\end{array}\right)\left(\begin{array}{c}E_x\\E_y\\E_z\end{array}\right)=0
\end{align}
The roots of the determinant of the previous matrix provides the eigenmodes of propagation. The following system has to be solved:
\begin{align}
\left|\begin{array}{ccc}
S-n^2\cos^2\theta&-iD&n^2\sin\theta\cos\theta\\
iD&S-n^2&0\\
n^2\sin\theta\cos\theta&0&P-n^2\sin^2\theta
\end{array}\right|=0
\end{align}
The general solution of this equation is known as the Appleton-Hartree equation:
\begin{align}
n^2=1-\frac{2X(1-X)}{2(1-X)-Y^2\sin^2\theta\pm\sqrt{Y^4\sin^4\theta+4(1-X)^2Y^2\cos^2\theta}}
\end{align}
This equation links the scalar refractive index $n$ with $X(\omega)$, $Y(\omega)$ and $\theta$.

\begin{figure}
\centering\includegraphics[width=0.8\textwidth]{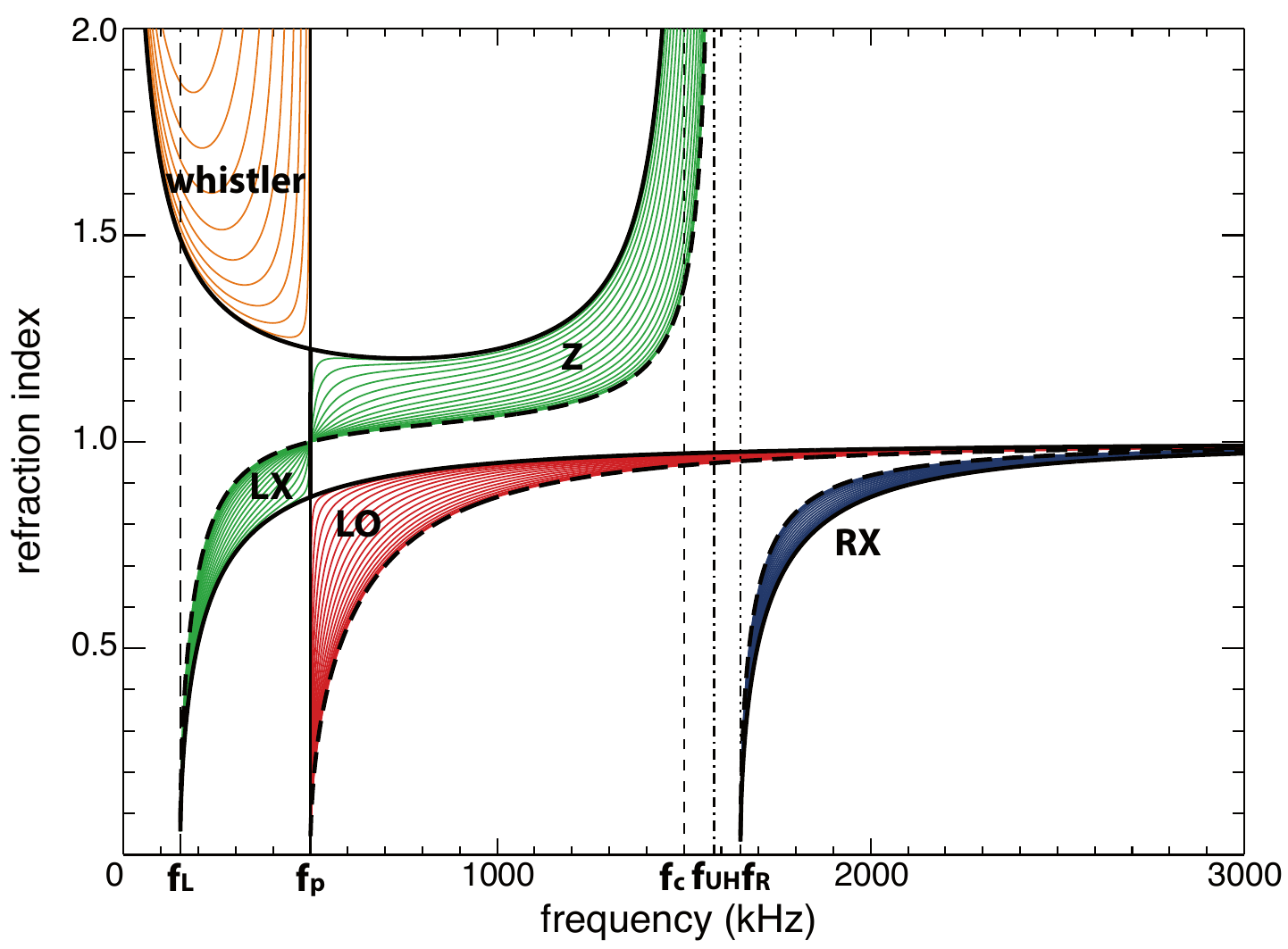}
\caption{Dispersion diagram with following plasma and cyclotron frequency values: $f_p=500$\,kHz and $f_c=1500$\,kHz. Vertical lines, from left to right are: $f_L$, L-mode cutoff frequency; $f_p$, plasma cutoff frequency; $f_c$, cyclotron resonance frequency; $f_{UH}$ upper hybrid resonance frequency; and $f_R$, R-mode cutoff frequency.}\label{fig:dispersion}
\end{figure}

\subsection{Propagation Modes, propagation angle}
The $\pm$ signs in the Appleton-Hartree equation correspond to two eigenmodes and thus two propagation modes. We hereafter refer to them as the $\oplus$ and $\ominus$ modes, respectively. Figure (\ref{fig:dispersion}) shows the various propagations modes. Each curve on the figure correspond to a propagation angle with respect to the ambiant magnetic field $\bm{B}_0$: from parallel propagation in bold plain line; to perpendicular propagation in bold dashed line. Each region of the diagram is a specific propagation mode and is named as shown on the figure. The $\oplus$ modes are the {\it whistler} and {\it LO} modes, whereas the $\ominus$ modes are the {\it Z}, {\it LX} and {\it RX} modes. The vertical lines are the {\it resonnance} (when $n\rightarrow\infty$) and {\it cutoff} (when $n\rightarrow 0$) frequencies.

\subsection{Polarization}
When studying wave propagation in a cold plasma, the sense of polarization is always given considering the wave propagating in the direction of the magnetic field. 

In case of parallel propagation, the polarization of the wave is always circular. The $\oplus$ and $\ominus$ modes are respectively fully LH and RH polarized. 

In case of perpendicular propagation, the $\oplus$ mode is linearly polarized along the ambiant magnetic field direction, and the $\ominus$ mode is elliptically polarized perpendicular to the ambiant magnetic field direction (this means that the wave propagation is not transverse). 

In case of oblique propagation, the $\oplus$ and $\ominus$ modes are elliptically polarized.

\section{Radio Waves Propagation}
The propagation of radio waves is usually studied with ray tracing methods. They can include scattering effects. The main ray tracing algorithm for radio-astronomy is the Haselgrove algorithm \cite{haselgrove_55,haselgrove_63}. Other algorithms are available (such as Poeverlein \cite{poverlein_48}), but are mainly adapted to horizontally stratified propagation media, such as the Earth's ionosphere. 

\begin{figure}
\centering\includegraphics[width=\textwidth]{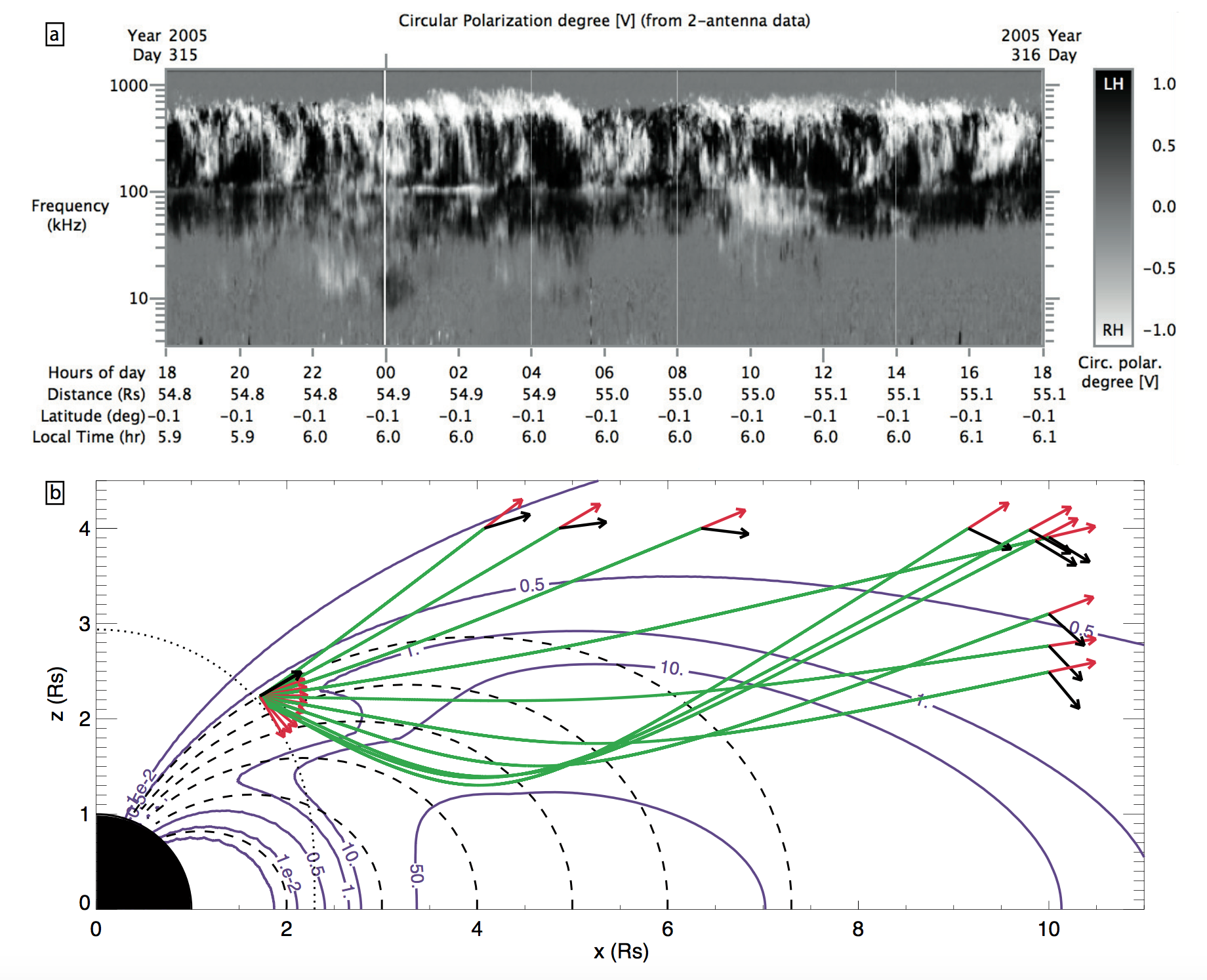}
\caption{Ray tracing in the inner Saturn magnetosphere. a) Dynamic spectrum of circular polarization degree of SKR. b Ray tracing performed by ARTEMIS-P through the Kronian magnetosphere at $f=50\,\textrm{kHz}$. Green lines are ray trajectories, red arrows are wave vectors, black arrows are directions of the magnetic field, blue line are electron density contours at $[0.001,0.01,0.5,1,10,50] \textrm{cm}^{-3}$, dotted black line is the iso-surface where $f_{ce} = 50\,\textrm{kHz}$, long dashed black lines are dipolar magnetic field lines. The magnetospheric plasma density is taken from the MDISC model \cite{achilleos_2010}. Figure adapted from \cite{gautier_13}}\label{fig:ex1:saturn-magnetosphere}
\end{figure}

The two main propagation effects are the refraction and the scattering of radio waves. While the former is directly related to the evolution of the refractive index along the ray path, the latter includes random changes in ray propagation, due to interaction is particle and inhomogeneities in the propagation medium. Refraction is the core of the ray tracing techniques. Scattering can be included \cite{steinberg_04}. Further non-linear effects, such as caustics can not be handled by such simulation codes. Full electromagnetic code (such as Finite Difference Time Domain (FDTD) methods) should then be used.

Figure (\ref{fig:ex1:saturn-magnetosphere}) 
is showing an examples the effect of radio wave propagation. It displays the trace of several radio rays in the inner magnetosphere of Saturn, at a frequency of $50\,\textrm{kHz}$. The apparent location of the radio source as seen from a remote location highly depends on the initial ray inclination with the source ambient magnetic field vector. 

\subsection{Propagation of polarization}
The propagation of polarization can be treated in ray tracing codes. Two kind of propagation regions shall be separated, where (1) the medium is imposing the polarization on the wave, or (2) the wave is freely propagating with its own polarization. These regions are called ``weak'' and ``strong'' mode coupling, respectively. The ``coupling'' is here referring to that between the wave modes, not between the waves and the medium. The limit between the two regimes is given by the following relation \cite{booker_36,budden_52}:
\begin{align}
|n_\oplus - n_\ominus | \sim \frac{1}{k}\frac{\mathrm{d}}{\mathrm{d}k}n_{(\oplus\textrm{ or }\ominus)}
\end{align}
In the strong mode coupling region, the polarization is frozen and is projected on the two modes of the medium, as in a birefringent medium in classic optics. If the two modes have different phase velocities, Faraday rotation occurs (see \cite{oberoi_2012} and references therein). 

\section{Detection}
\begin{figure}
\includegraphics[width=0.9\textwidth]{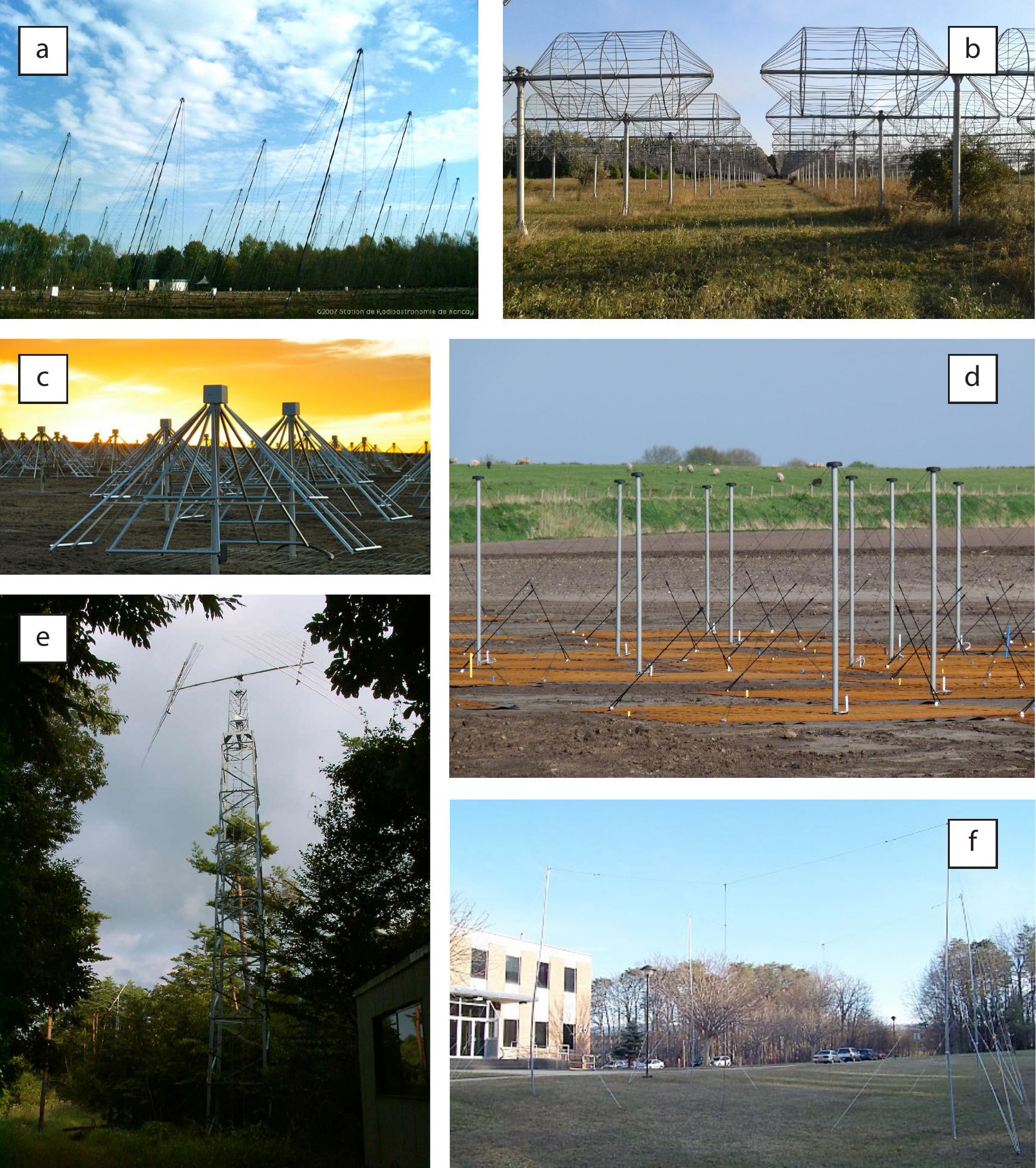}
\caption{Various types of ground based radio antenna used in the low frequency range (10\,MHz to 80\,MHz): (a) Array of helicoidal antenna, Nan\c cay Decameter Array (Nan\c cay, France); (b) Array of thick linear dipoles, UTR-2 (Kharkov, Ukraine); (c) Array of thick folded crossed dipoles, LWA (New Mexico, USA) and Nenufar (Nan\c cay, France); (d) Array of thin folded crossed dipoles, LOFAR-LBA (EU); (e) Log Periodic Yagi antenna, Iitate HF Radio Monitor (Iitate, Japan); (f) thin dipole, RadioJOVE (world wide).}\label{fig:ground-radio}
\end{figure}

The Earth ionosphere is reflecting radio waves below 10\,MHz whether the wave is coming from space or from ground. This is used for telecommunication on ground in the so-called long wavelengths radio bands. Space based observations are required below 10\,MHz. Between 10\,MHz and $\sim$80\,MHz (i.e., the between ionospheric cutoff and the FM broadcasting band), several ground based instruments are available. Figure (\ref{fig:ground-radio}) is showing several types of radio antenna used for ground based observations. 

The angular resolution of a telescope is defined by the ratio between the observed wavelength and the telescope aperture diameter. Hence, in order to get a few arcseconds resolution at 30\,MHz (i.e., a wavelength of 10\,m), an aperture of about 2500\,km in needed. As single piece focussing devices of this size cannot be built on ground, phased array are used together with interferometric techniques.

\subsection{Space based sensors}
In the vicinity of Earth, the accessible spectral range includes frequencies as low as a few kHz for observations of radio waves above the local propagation cut-off. The constraints on space based instrumentation, such as reliability and cost imply that only simple sensors can be used on a spacecraft. Space-based electric and magnetic sensors are very simple. Electric sensors are either wire boom monopoles or dipoles, or a pair of probe antenna. Magnetic sensors are either magnetic loops (single loop) and search coils (many loops). We will focus in the following on wire dipole electric sensors. 

Space based low frequency radio interferometry has not been implemented yet, for obvious reasons of cost for building and operating a fleet of several tens or hundreds of spacecraft. Nanosatellite concepts may help, and several studies are ongoing \cite{oberoi_2005,rajan_11,burns_11}.

\subsection{Presentation of Goniopolarimetry}
\begin{figure}
\includegraphics[width=\textwidth]{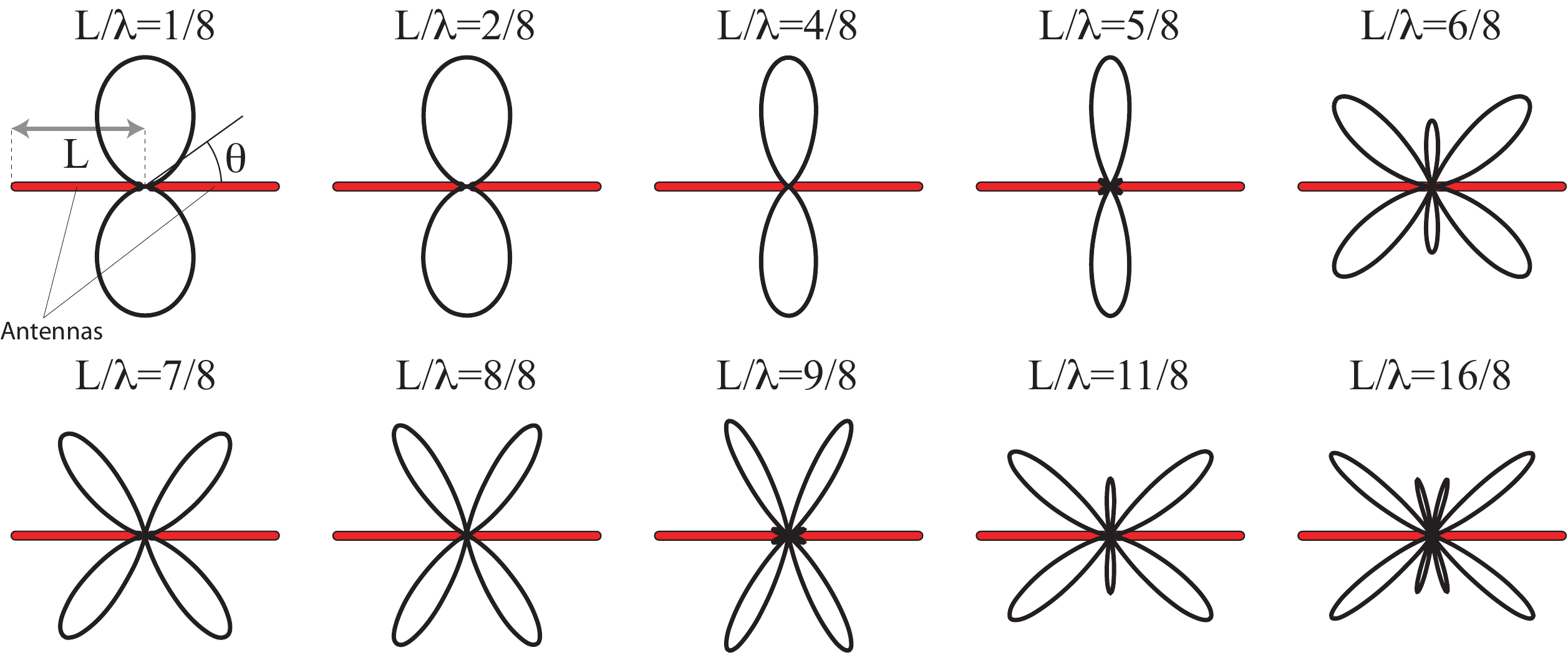}
\caption{Linear dipole antenna beaming pattern for various wavelengths to antenna lengths ratio.}\label{fig:dipole-pattern}
\end{figure}

At frequencies much lower than the resonance frequency of wire dipole electric sensors (i.e., at wavelengths longer than 20 times the antenna length, so-called {\it short antenna} or {\it quasistatic} range), the gain properties of the antenna is very simple: 
\begin{align}
G(\theta) \propto \sin^2\theta,
\label{eq:short-antenna-gain}
\end{align}
where $\theta$ is the angle between the antenna axis and the wave vector direction $\bm{k}$. Figure (\ref{fig:dipole-pattern}) is showing the antenna beaming pattern at different wavelengths. In the quasistatic range (upper left corner), the beaming pattern is single lobed, and its gain figure varies as presented in Eq.\ (\ref{eq:short-antenna-gain}). When the wavelength is getting shorter, the beaming pattern is getting narrower, up to the resonance frequency ($L/\lambda=1/2$) and then becomes multi-lobed. 

\begin{figure}
\centering\includegraphics[width=0.7\textwidth]{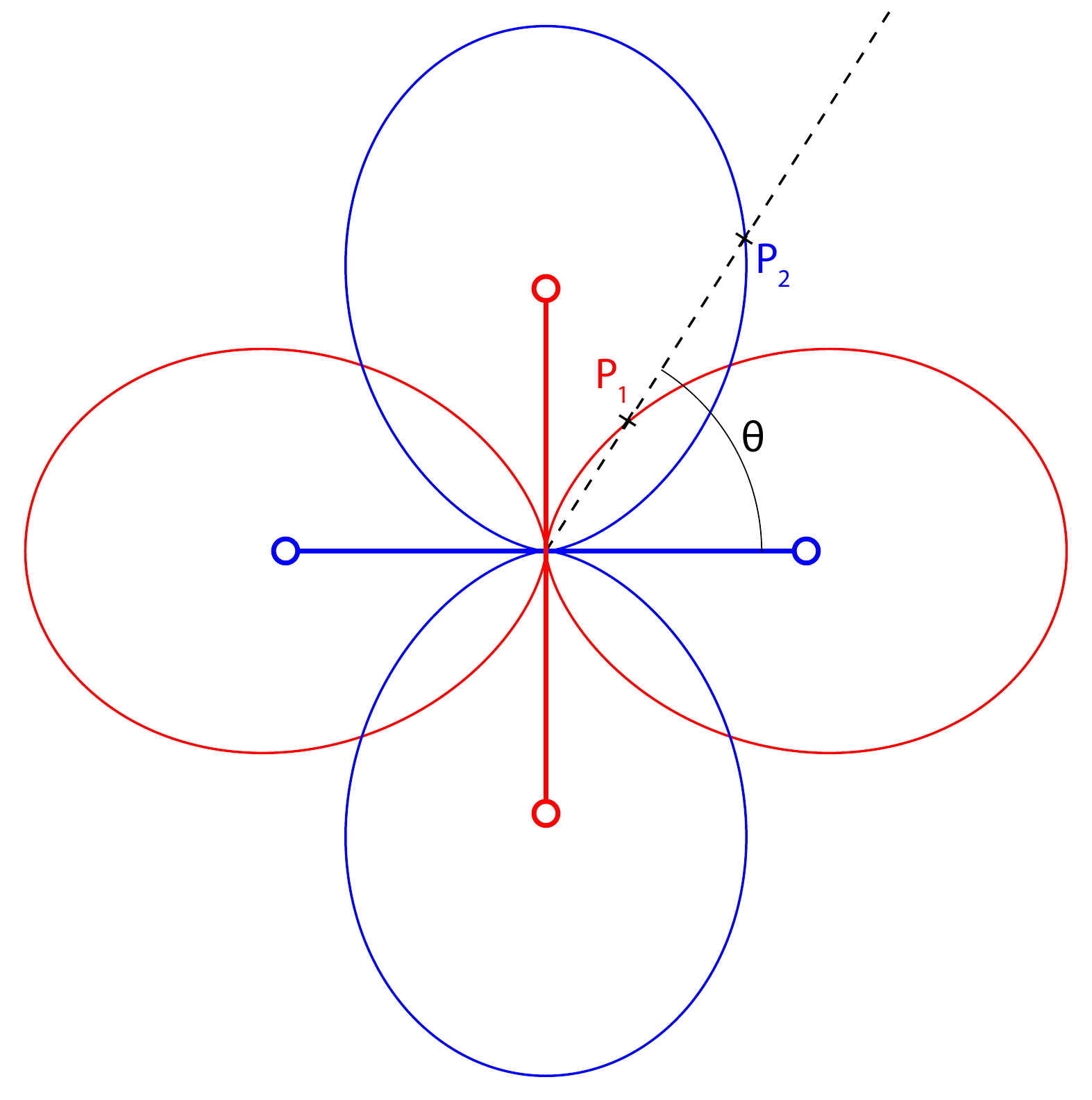}
\caption{Principle of goniopolarimetry illustrated in a simple planar case. The wave arrives with an angle $\theta$ from the blue antenna. The ratio of the radio power measured on each antenna is directly linked to $\theta$.}\label{fig:simple-gonio}
\end{figure}

Goniopolarimetric (i.e., simultaneous derivation of {\it direction of arrival} and {\it polarization}) inversions are making use of the simple form of the beaming pattern in the quasistatic range. Figure (\ref{fig:simple-gonio}) is showing a  simplified case of linearly polarized wave observed with a set of orthogonal crossed dipoles, in a fully planar configuration. The direction of arrival $\theta$ can simply be determined after the ratio of the $P_1$ and $P_2$ measured power. The solution is not unique in this case. 

For a three-dimensional analysis, either three quasi-orthogonal dipoles, or dipoles on a spinning spacecraft (1 aligned with the spin axis, the other(s) in the spin plane) shall be used. In the quasistatic range, the electric signal induced by the wave on an antenna is:
\begin{align}
V_a=\bm{h}_a.\bm{E}_\omega(t),
\end{align}
where $\bm{h}_a$ is the antenna vector and $\bm{E}_\omega(t)$ the wave instantaneous electric field. The radio wave is assumed to be in a transverse propagation regime. Hence, the instantaneous electric field is: 
\begin{align}
\bm{E}_\omega(t) = e^{i(\omega t + \phi_0)}\left|\begin{array}{l}
a_x\\a_ye^{i\delta}\\
0\end{array}\right.
\end{align}
in the wave polarization plane (i.e., perpendicular to $\bm{k}$). 

Goniopolarimetric radio receivers are measuring the power on each sensor, as well as the cross-correlation of the signals measured on each pair of antennas. This provides us with the full spectral matrix $P_{ij}$ of the wave \cite{ladreiter_95,cecconi_2005,cecconi_2007}:
\begin{align}
P_{ij} = \frac{Z_0Gh_ih_jS}{2}&\left[(1+Q)A_iA_j + (U-iV)A_iB_j\right.\nonumber\\
&\left.+ (U+iV)A_jB_i + (1-Q)B_iB_j\right] 
\end{align}
where: $S$, $Q$, $U$ and $V$ are the four Stokes parameters of the wave; $h_i$ and $h_j$ the antenna electrical lengths; $G$, the gain of the receiving chain; $Z_0$ the impedance of free space; and $A_n$ and $B_n$ (with $n=i$ or $n=j$) defined as:
\begin{align}
A_n &= -\sin\theta_n\cos\theta\cos(\phi-\phi_n)+\cos\theta_n\cos\theta\\
B_n &= -\sin\theta_n\sin(\phi-\phi_n)
\end{align}
where: $\theta$ and $\phi$ defines the direction of the wave vector $\bm{k}$ in a spherical frame; and $\theta_n$ and $\phi_n$ defines the direction of the $n$-th antenna $\bm{h}_n$ in the same frame. All the spectral matrix terms $P_{ij}$ are not always measured, depending on the receiver capabilities. Goniopolarimetric techniques are inverting the set of measurements $P_{ij}$ into the wave parameters. The Stokes parameters can be related to the wave parameters as follows: 
\begin{align}
S & =\frac{<E_x.E_x^*> + <E_y.E_y^*>}{2Z_0} = \frac{<a^2_x> + <a^2_y>}{2Z_0}\\
Q & =\frac{<E_x.E_x^*> - <E_y.E_y^*>}{<E_x.E_x^*> + <E_y.E_y^*>} = \frac{<a^2_x> - <a^2_y>}{2SZ_0}\\
U & =\frac{<E_x.E_y^*> + <E_y.E_x^*>}{<E_x.E_x^*> + <E_y.E_y^*>} = \frac{<a_xa_y\cos\delta>}{SZ_0}\\
V & =-\frac{<E_x.E_y^*> - <E_y.E_x^*>}{i(<E_x.E_x^*> + <E_y.E_y^*>)} = \frac{<a_xa_y\sin\delta>}{SZ_0}
\end{align}
The radio convention \cite{kraus_66} of the sign of $U$ and $V$ is the opposite of the optical convention. Radio astronomers consider they measure the polarization following the radio wave instead of looking at the source.

The goniopolarimetric inversions are used to derive the wave polarization state. The plane of polarization is also determined. The direction of propagation is then derived as the normal direction to the polarization plane, with the transverse propagation assumption. This assumption is valid most of the time, except close to the wave propagation resonances and cutoffs. In that case, determining the polarization of the magnetic component is directly providing the direction of $\bm{k}$. However, in such a case, the direction of propagation (the Poynting vector) is not that of the wave vector. Hence, simultaneous electric and magnetic component sensing is required. 

In case of planetary radio emissions, the radio source location (or at least the magnetic hemisphere containing its location) can be derived from the direction of arrival, see Figure (\ref{fig:radio-source-loc}), and the sense of polarization indicates the emission mode. In weak mode coupling conditions (the medium is setting the polarization), the sign of the Stokes parameter $Q$ is providing the propagation mode:
\begin{itemize}
\item $Q>0$ means $\oplus$ mode (L-O branch of the dispersion diagram)
\item $Q<0$ means $\ominus$ mode (R-X branch of the dispersion diagram)
\end{itemize}

\subsection{Various goniopolarimetric inversions}
Several goniopolarimetric inversions are available in the literature. The following list is grouping them with their assumptions:
\begin{itemize}
\item {\it Point sources}. The output parameters are $S$, $Q$, $U$, $V$, $\theta$ and $\phi$. They are used for auroral radio emissions (Earth, Jupiter, Saturn). They have been used on Cassini/RPWS (with 2 or 3 antenna modes) and INTERBAL/Polrad (with 3 antennas) \cite{cecconi_2005,ladreiter_95,lecacheux_78,panchenko_2003}.
\item {\it Spatially extended sources}. The output parameters are $S$, $Q$, $U$, $V$, $\theta$, $\phi$ and $\gamma$, where $\gamma$ is the angular size of the source (assuming a circular shaped source region). They are used for solar radio emissions. They have been used on STEREO/Waves (with 3 antennas) and WIND/Waves (spinning spacecraft) \cite{manning_80,cecconi_2007,krupar_2012}.
\item {\it Radio sources along a spatial profile}. The output parameters are $S(a)$, $Q(a)$, $U(a)$, $V(a)$, with $a$ the curvilinear coordinate along the spatial profile. Such inversion can be used for auroral radio sources \cite{hess_2010}.
\item {\it All sky source}. The output parameters are the Stokes parameters distribution on the sky $S(\theta,\phi)$, $Q(\theta,\phi)$, $U(\theta,\phi)$, $V(\theta,\phi)$. Such inversions can be used for mapping the galactic background emission, but are still to be developed. The only Galactic emission mapping below $10\,\textrm{Mhz}$ was obtained with Lunar occultations measured by Radio Astronomy Explorer 2 (RAE 2) \cite{novaco_78}.
\end{itemize}

\subsection{Limitations}
Although very powerful compared to the use of a single dipole, goniopolarimetric system have several intrinsic limitations:
\begin{itemize}
\item {\it Electric versus magnetic sensors}. Magnetic component intensity is $c$ times fainter than the electric component. Furthermore, the magnetic sensor have an intrinsic band pass input filter that forbids wide band applications.
\item {\it Sense of propagation}. Simple goniopolarimetry provides the wave direction of propagation, but not its sense of propagation. Coupled electric and magnetic components sensing are required to do so. 
\item {\it Wave vector direction}. Goniopolarimetric inversions with electric sensors assume transverse propagation. It is not true when $n\neq 1$. As discussed earlier, observing the magnetic components provides directly the wave vector direction. However, the direction of propagation is not that of the wave vector. 
\item {\it Accuracy}. The goniopolarimetric accuracy depends on two factors: the accuracy of the receiver system calibration (effective antenna directions and lengths, phase miss-match between sensing channels); and the receiver and onboard processing noise. The main source of noise is usually the digitization noise, which quantifies the output signal. This results in the following typical accuracies: $\sim$2$^\circ$ on direction of arrival, $\sim$10\% on polarization degrees and $\sim$3\,dB on flux densities. 
\item {\it Effective sensor directions}. Electric and magnetic sensors must be accurately calibrated to obtain their effective parameters (length and direction). This calibration process is more critical for electric sensors, as the conductive spacecraft body can strongly alter the antenna beaming pattern.
\item {\it Propagation mode}. The propagation mode can not be derived unambiguously without simultaneous and coupled measurements on electric and magnetic sensors.
\item {\it Direction of propagation}. The wave vector is aligned with the Poynting vector, except when $n\neq 1$.
\item {\it Radio source location}. The goniopolarimetric systems are measuring the wave parameter at the place of the spacecraft. Hence the radio source location can not be retrieved without specific assumptions, such as: straight line propagation, emission process. 
\end{itemize}

\section{Example of results and open questions}
According to the magneto-ionic theory, there is a direct link between the propagation angle and the polarization. As seen in the previous section, perpendicular emission leads to linear or elliptical polarization, parallel emission to circular polarization, and oblique emission to elliptical polarization. The sense of polarization depends on the direction of the magnetic field in the source (hence on the hemisphere of emission), as well as on the propagation mode.

Solar radio emissions are showing a limited degree of polarization. In the lower frequency range (below $\sim$10\,MHz), radio observations are not showing evidences of significant degree of polarization (excepty for a limited series of event \cite{reiner_07}), whereas at higher frequency, up to 35\% of polarization can be observed \cite{dulk_80}. Contrarily planetary radio emissions are fully polarized, either circularly or elliptically \cite{zarka_1992}. The planetary radiation belts are elliptically or linearly polarized \cite{chang_62,vesecky_67,legg_68,levin_01}.

\subsection{Latitudinal variability of polarization}
\begin{figure}
\centering\includegraphics[width=0.46\textwidth]{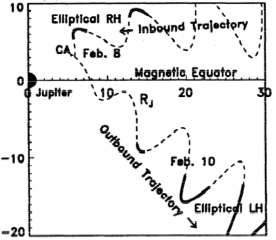}\includegraphics[width=0.5\textwidth]{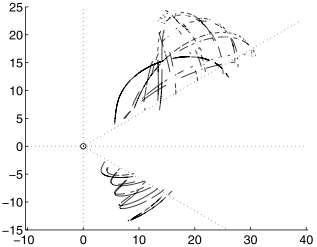}
\caption{Observed polarization state at Jupiter and Saturn, with Ulysses/URAP and Cassini/RPWS respectively. The left-hand panel shows a meridian plane projection of the trajectory of the Ulysses spacecraft during its Jovian flyby \cite{reiner_95}. Axes are number in units of Jovian Radii. Bold segments of the trajectory correspond to the location where elliptical polarization was observed. The right-hand panel shows a similar figure measured at Saturn \cite{fischer_2009}. Axes are number in units of Kronian Radii. The displayed fraction of orbit correspond to elliptical polarization waves. On both cases, circularly polarized waves are observed elsewhere.}\label{fig:observed-polar-latitude}
\end{figure}

All magnetized planets are producing radio emissions in their auroral regions. The main emission is produced in the R-X mode (i.e., RH polarized in the Northern magnetic hemisphere). The radio waves are 100\% polarized. At Earth, the observations show circularly polarized waves. At Jupiter and Saturn, auroral radio waves are circularly polarized when observed from near equatorial regions, while they are elliptically polarized at high latitudes. The limit latitude is about $\pm$30$^\circ$ as shown in Figure (\ref{fig:observed-polar-latitude}) \cite{reiner_95,fischer_2009}.  

\subsection{Saturn Kilometric Radiation}
\begin{figure}
\centering\includegraphics[width=\textwidth]{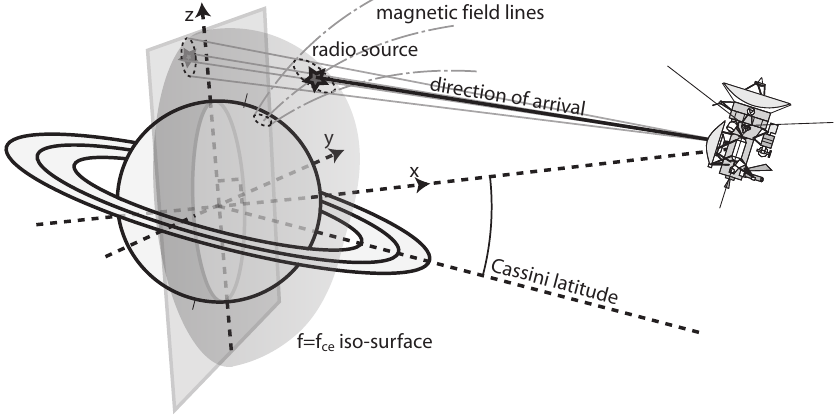}
\caption{Derivation of the radio source location at Saturn, from the goniopolarimetric products provided by Cassini/RPWS/HFR. Straight line propagation is assumed, together with CMI emission process, which implies an emission at the local electron cyclotron frequency ($f_{ce}$)in the source. The three-dimensional location of the radio source is derived, as well as the location of the magnetic footprint of the source (for comparison with atmospheric aurora). The error on the wave propagation direction is also used to derive the radio source footprint error ellipse. Figure extracted from \cite{cecconi_2009}.}\label{fig:radio-source-loc}
\end{figure}

\begin{figure}
\centering\includegraphics[width=\textwidth]{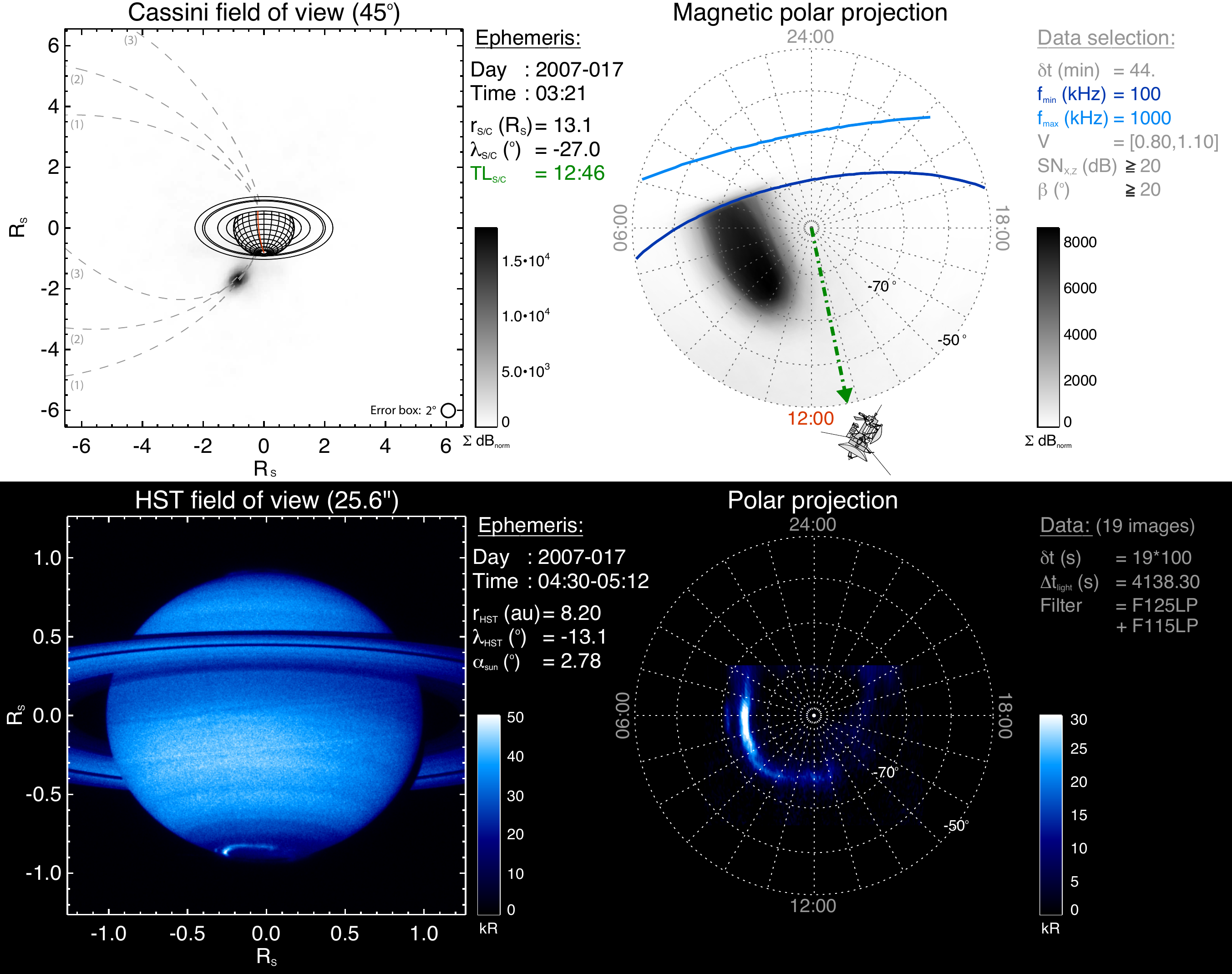}
\caption{Comparison of radio source location with UV aurora at Saturn. The top panel shows radio source locations derived from Cassini/RPWS/HFR measurements. The bottom panel shows the UV aurora as observed from HST. The left-hand column is displaying the radio and UV sources as seen from the observer, whereas the right-hand column shows the same data projected on the southern polar ionosphere. Figure extracted from \cite{lamy_2009}.}\label{fig:uv-radio}
\end{figure}

Cassini/RPWS/HFR (Radio and Plasma Waves Science/High Frequency receiver) is a goniopolarimetric radio receiver. Cecconi et al.\ \cite{cecconi_2009} proposed a simple way to derive the radio sources locations assuming straight line propagation and CMI emission, as shown in figure (\ref{fig:radio-source-loc}). Figure (\ref{fig:uv-radio}) shows a comparison of active radio source magnetic footprints with ultraviolet (UV) aurora observed with the Hubble Space Telescope (HST), at Saturn \cite{lamy_2009}. Both datasets were observed at the same time, and with a very similar observing geometry. The figure shows an very good conjugacy between the two phenomena, which comforts the idea that the same electron populations are producing radio and UV emissions.

\begin{figure}
\centering\includegraphics[width=\textwidth]{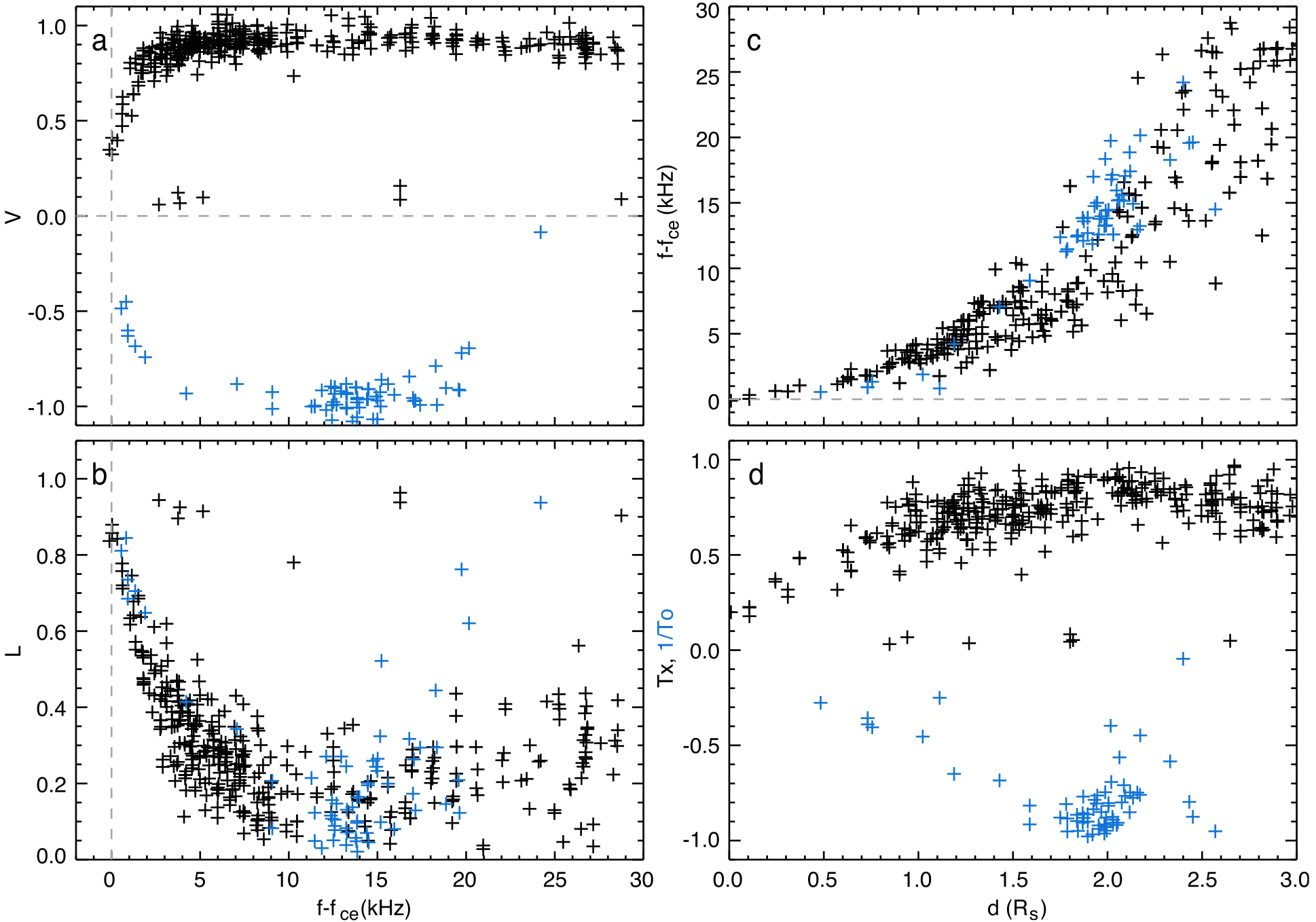}
\caption{Polarization parameters of Saturn Kilometric Radiation in the vicinity of a source region: (a) Circular polarization degree plotted versus the reduced frequency Figure extracted from \cite{lamy_2011}.}\label{fig:skr-source-polar}
\end{figure}

On Oct.\ 17th 2008, the Cassini spacecraft flew through the auroral radio source region for a few minutes \cite{lamy_2011}. Figure (\ref{fig:skr-source-polar}) shows the polarization parameters with respect to the distance to the radio source. It confirms that the wave has a quasi perpendicular propagation, with elliptical polarization consistently with the magnetoionic theory.

\subsection{Future prospects}

Goniopolarimetric methods are using measurements acquired on a single location. Although implying several assumptions they are very efficient for auroral planetary radio emissions, which are point sources with a high degree of intrinsic polarization. Solar radio bursts can also be measured, but a limited view of their spatial structure can be derived. 

The two main limitations of goniopolarimetric inversions are the assumptions of (i) a plane transverse wave propagating on a straight line from the source to the observer and (ii) a point or circularly-shaped single source observed at a time. The first limitation can be fixed using sensors measuring the full electric and magnetic components of the wave. This set up is proposed for the Alfv\'en space mission \cite{berthomier_11}, which is dedicated to the study of the Terrestrial auroral cavities. Solving the second limitation implies space-based radioastronomy interferometric instrumentation \cite{oberoi_2005,burns_11}.

\end{document}